\documentclass[aps,10pt,notitlepage,twocolumn,prd,preprintnumbers,longbibliography,nofootinbib,showpacs]{revtex4-2}
\usepackage[colorlinks=true,urlcolor=blue,linkcolor=blue,citecolor=blue]{hyperref}
\usepackage{slashed,color,amsmath,amssymb,mathrsfs}
\usepackage{amsfonts,epsfig,amssymb}
\usepackage{graphicx}
\usepackage{float}
\usepackage{longtable}
\usepackage{array}
\usepackage{accents}
\usepackage{tensor}
\usepackage{footmisc}
\usepackage{longtable}
\usepackage{booktabs}
\usepackage{multirow}
\usepackage{bm}
\usepackage{subfigure}
\usepackage{gensymb}
\usepackage{textcomp}
\usepackage{dcolumn}
\usepackage{tikz}
\usepackage{bm}

\allowdisplaybreaks

\begin{document}
\title{\texorpdfstring%
{An anlaysis on $J/\psi\to\pi^0\gamma^*$ within resonance chiral theory}%
{An anlaysis on Jpsitopigamma within resonance chiral theory}}
	\author{Yi-Hao Zhang$^{1,2,3}$}
    \author{Shao-Zhou Jiang$^{1}$}
	\author{Ling-Yun Dai$^{2,3}$}
	\email{dailingyun@hnu.edu.cn}
\affiliation{$^{1}$ Key Laboratory for Relativistic Astrophysics, School of Physical Science and Technology, Guangxi University, Nanning 530004, People’s Republic of China}
	\affiliation{$^{2}$ School for Theoretical Physics, School of Physics and Electronics, Hunan University, Changsha 410082, China}
	\affiliation{$^{3}$ Hunan Provincial Key Laboratory of High-Energy Scale Physics and Applications, Hunan University, Changsha 410082, People's Republic of China}

\date{\today}

\begin{abstract}
In this study, we analyze the first measurement of the electron-positron invariant mass spectrum in $J/\psi \to \pi^0 e^+e^-$ by BESIII, using the framework of resonance chiral theory. Our results indicate that both strong interaction and electromagnetic transition are essential to accurately describe the data.
We obtain the $\pi^0$ transition form factor for $J/\psi \to \pi^0\gamma^*$ and the corresponding decay branching ratios for $J/\psi \to \pi^0 l^+l^-$. The decay process $J/\psi \to \pi^0 V$ is also examined. It is found that $J/\psi \to \pi^0 \rho^0$ is dominated by the strong interaction, while the other two channels, $J/\psi \to \pi^0 \omega$ and $\pi^0 \phi$, arise primarily from electromagnetic transitions.
\end{abstract}

	\maketitle


\section{Introduction}
Hadronic decays of $J/\psi$ provide an ideal laboratory for testing Quantum Chromodynamics (QCD)~\cite{Greiner:2007zz}. The BESIII experiment has already collected more than 10 billion $J/\psi$ events \cite{BESIII:2021cxx}.
Among these, $J/\psi\to \pi^0\gamma$ and $J/\psi\to \pi^0e^+e^-$ are two of the lightest decay modes containing at least one hadron. Although rare, these decays serve as sensitive probes for investigating the inner structure of hadrons and offer insights into the non-perturbative regime of QCD in this energy region.
Recently, BESIII reported the first measurement of the $e^+e^-$ invariant mass spectrum for the process $J/\psi \to \pi^0 e^+e^-$~\cite{BESIII:2025xjh}. The result clearly reveals a $\rho$-$\omega$ interference effect and a dip structure in the $\phi$ mass region. Furthermore, the data above 1.2~GeV provide valuable information on heavier vector resonances. In Ref.~\cite{BESIII:2025xjh}, several preliminary theoretical predictions, including those from dispersion relations~\cite{Kubis:2014gka,JPAC:2023nhq} and resonance chiral effective theory (RChT)~\cite{Chen:2014yta,Yan:2023nqz}, were compared with the data, but none provided a fully satisfactory description. Nevertheless, previous studies~\cite{Chen:2014yta,Yan:2023nqz,Fu:2011yy} have shown that RChT and vector meson dominance (VMD) work well for similar processes such as $J/\psi\to \eta^{(')}\gamma^*$. Therefore, we perform a comprehensive analysis of $J/\psi\to \pi^0\gamma^*$ based on RChT to examine whether it can improve the description of the experimental measurements reported in Ref.~\cite{BESIII:2025xjh}.

In the present analysis, the decay widths of $J/\psi\to\pi^0 V$, with $V$ being the lightest vector mesons $\rho$, $\omega$, and $\phi$, can be studied simultaneously. The long-standing puzzle of $\psi(nS)\to\rho\pi$, dating back to the 1980s~\cite{Franklin:1983ve}, has presented a major theoretical challenge in understanding charmonium decays for decades~\cite{Chen:1998ma,Suzuki:2000yq,Brodsky:1997fj,Pinsky:1989ue,Zhao:2010ja}. Recent works attribute this to a destructive interference between the strong interaction and electromagnetic (EM) transition  amplitudes~\cite{Yan:2023nqz,Kivel:2023fgu}. Our analysis aims to quantitatively clarify the contribution from each part by considering all the three decay channels, $J/\psi\to\pi^0 \rho$, $\pi^0 \omega$, and $\pi^0 \phi$.
The electromagnetic transition is computed using the pion transition form factor (TFF) from our previous work~\cite{Zhang:2025ijd}, which yields significant effects in the $\rho$–$\omega$ and $\phi$ energy regions. Correspondingly, the new data could offer valuable insight into the time-like doubly-virtual TFF $\mathcal{F}_{\pi^0 \gamma^* \gamma^{*}}(M_\psi^2,q^2)$, particularly around $q^2 \approx M_{\rho,\omega, \phi}^2$. It is found that the observed structure around $\rho$–$\omega$ energy region is well described as a coherent superposition of distinct production mechanisms: the $\rho$ resonance originates from the strong interaction (leading-order (LO) amplitude), while the $\omega$ resonance is introduced via the electromagnetic transition. This mechanism produces a significantly larger effect than the $\rho$–$\omega$ mixing included within the LO strong interaction amplitude alone. Furthermore, we find that charmonium contribution must be taken into account to properly describe the data. Although smaller than the LO strong interaction amplitude, its contribution becomes larger than the EM transition in the higher-energy region above 1.5 GeV.

This paper is organized as follows. In Section~\ref{sec2}, we introduce the theoretical framework of RChT and derive the TFF for $J/\psi \to \pi^0 \gamma^*$. In Section~\ref{sec3}, we present the analysis of the data using the aforementioned framework, including predictions for the decay widths of $J/\psi\to \pi^0 V$ and the $\mu^+\mu^-$ invariant mass spectrum of $J/\psi\to\pi^0\mu^+\mu^-$.
Finally, a summary of our conclusions is provided in Section~\ref{sec5}.

\section{Theoretical framework}
\label{sec2}
\subsection{Construction of the RChT Lagrangian}
We employ chiral effective field theory to study the interactions among hadrons in the relevant energy region. In chiral perturbation theory (ChPT) \cite{Weinberg:1978kz,Gasser:1983yg}, the lowest-lying pseudoscalar octet is treated as the Goldstone bosons arising from the spontaneous breaking of chiral symmetry for the light quarks ($u$, $d$, and $s$). ChPT successfully describes the low-energy interactions of these pseudoscalars. At the higher energies, resonances such as $\rho$, $\omega$, and $\phi$ emerge, and their interactions with the pseudoscalars can be described by RChT \cite{Ecker:1988te,Ecker:1989yg,Cirigliano:2006hb,Kampf:2006yf,Portoles:2010yt,Kampf:2011ty}. RChT extends ChPT by introducing these resonances as explicit degrees of freedom.
The effective Lagrangian used in our analysis is taken from Refs.~\cite{Zhou:2025cev,Qin:2020udp,Wang:2023njt,Zhang:2025ijd,Chen:2012vw,Chen:2014yta,Yan:2023nqz},
\begin{eqnarray}
\mathcal{L}&=&\mathcal{L}^{V} _{\mathrm{kin}}+\mathcal{L} _{\mathrm{WZW} }+\mathcal{L} _{VJ}  +\mathcal{L}_{VJ P}+\mathcal{L}_{VVP}
\nonumber\\
&+&\mathcal{L} _{\psi  J}+\mathcal{L}_{\psi J P}+\mathcal{L}_{\psi V P}\,, \label{Eq:L;int}
\end{eqnarray}
where the subscripts ``$P$, $V$, $\psi $, $J$" represent the pseudoscalar mesons, the vector mesons, $J/\psi$, and electromagnetic current, respectively.

The Lagrangians in the first line, comprising the kinetic term of the vector mesons $\mathcal{L} _{\mathrm{kin} }^{V}$, the Wess-Zumino-Witten (WZW) term $\mathcal{L} _{\mathrm{WZW} }$ \cite{Witten:1983tw,Wess:1971yu}, the $V-\gamma$ coupling term $\mathcal{L} _{VJ}$, and the $VJP$, $VVP$ interaction terms, can be found in our previous work \cite{Zhang:2025ijd}.
Those in the second line are relevant to $J/\psi$ and have already been defined within the framework of $U(3)$ RChT in Refs.~\cite{Chen:2014yta,Yan:2023nqz}. The Lagrangian describing the $J/\psi-\gamma$ coupling is defined as,
\begin{equation}
    \mathcal{L} _{\psi J}=\frac{\sqrt{2}ef_{\psi  }}{3 M_{\psi  }}\ \hat{\psi }  _{\mu \nu }F^{\mu \nu } ,
\end{equation}
where $F^{\mu \nu}=\partial^\mu A^\nu-\partial^\nu A^\mu$ and $\hat{\psi } _{\mu \nu }=\partial _{\mu } \psi _{\nu } -\partial _{\nu } \psi _{\mu } $ denote the field strength tensors for the photon field $A_\mu$ and the $J/\psi$ field $\psi_\mu$, respectively. With this term, one can obtain the decay width
\begin{equation}
\Gamma_{\psi \to e^+e^-}=\frac{32e^4 f_{\psi }^2(2m_e^2+M_{\psi }^2)}{27 \pi M_{\psi }^3}.
\end{equation}
The parameter $f_\psi$ can be fixed by comapring our result with the PDG value~\cite{ParticleDataGroup:2024cfk}, $\mathrm{Br}_{J/\psi\to e^+e^-}^{\mathrm{PDG}}=(5.971\pm0.032)\%$, yielding $f_\psi= (0.294\pm 0.001)$~GeV.

The three-hadron interaction terms in Eq.~(\ref{Eq:L;int}) are given by Refs.~\cite{Chen:2014yta,Yan:2023nqz},
\begin{eqnarray}
\mathcal{L}_{\psi  J P}&=& \sum_{j=1}^{2} g^{\psi }_{j}\mathcal{\tilde{O}} _{{\psi  JP}} ^{j} \nonumber, \\
\mathcal{L}_{\psi(') VP}&=&\sum_{i=1}^{3} h^{\psi } _{i} \mathcal{\tilde{O}} _{{\psi  VP}} ^{i}\,,
\end{eqnarray}
where the $\psi  JP$ and $\psi  VP$ operators are given as
\begin{eqnarray}
    \mathcal{\tilde{O}}_{\psi  JP}^1 & = & \varepsilon_{\mu \nu \rho \sigma} \psi  ^\mu\left\langle\tilde{u}^\nu \tilde{f}_{+}^{\rho \sigma}\right\rangle, \nonumber \\
    \mathcal{\tilde{O}}_{\psi   JP}^2 & = & \frac{1}{M_{\psi  }^2}  \varepsilon_{\mu \nu \rho \sigma} \psi  ^\mu\left\langle\left\{\tilde{u}^\nu, \tilde{f}_{+}^{\rho \sigma}\right\} \tilde{\chi}_{+}\right\rangle, \\
    \mathcal{\tilde{O}}_{\psi   V P}^1 & = & M_{\psi  }  \varepsilon_{\mu \nu \rho \sigma} \psi  ^\mu\left\langle\tilde{u}^\nu V^{\rho \sigma}\right\rangle, \nonumber \\
    \mathcal{\tilde{O}}_{\psi   V P}^2 & = & \frac{1}{M_{\psi  }} \varepsilon_{\mu \nu \rho \sigma} \psi  ^\mu\left\langle\left\{\tilde{u}^\nu, V^{\rho \sigma}\right\} \tilde{\chi}_{+}\right\rangle, \nonumber \\
    \mathcal{\tilde{O}}_{\psi   V P}^3 & = & M_{\psi  } \varepsilon_{\mu \nu \rho \sigma} \psi  ^\mu\left\langle\tilde{u}^\nu\right\rangle\left\langle V^{\rho \sigma}\right\rangle.
\end{eqnarray}

In order to include $\rho-\omega$ mixing, we apply the momentum-dependent mixing mechanism given in Refs.~\cite{Gasser:1982ap,Wang:2023njt},
    \begin{equation}
        \binom{\left|\bar{\rho}^{0}\right\rangle}{|\bar{\omega}\rangle} =\left(\begin{array}{cc}
            \cos \delta & -\sin \delta_\omega(q^2) \\
            \sin  \delta_\rho(q^2) & \cos \delta
        \end{array}\right)\binom{\left|\rho^0\right\rangle}{|\omega\rangle} .
    \end{equation}
where $\bar{\rho}^0$ and $\bar{\omega}$ are the physical states and  $\delta=-1.8^\circ$~\cite{Wang:2023njt,Zhang:2025ijd} is the $\rho-\omega$ mixing angle, and the non-diagnonal parts are given as
\begin{eqnarray}
    \sin \delta_\omega (q^2) & = & -\sin\delta \frac{M_V\Gamma _V(q^2)}{\Delta^\ast_V(q^2) }, \nonumber \\
    \sin \delta_\rho (q^2) & = & \sin\delta \frac{M_V\Gamma_V(q^2)}{\Delta_V(q^2)  }.
\end{eqnarray}
Here, $\Delta_V(x) = M_V^2 - x - i M_V \Gamma_V(x)$ is the denominator of the Breit-Wigner (BW) propagator. One can take $V=\rho$ for simplicity. The width $\Gamma_\rho(x)$ includes a step function to ensure it vanishes when the lowest threshold of its decay channels is not open.
The $\omega$-$\phi$ mixing mechanism is adopted from Refs.~\cite{Dai:2013joa,Qin:2020udp}.
\begin{eqnarray}
\label{eq:V;thetaV}
\left ( \begin{array}{c}
\omega_{8} \\
\omega_{0}
\end{array}
\right ) &= \left ( \begin{array}{cc}
\cos\theta_{V} & \sin\theta_{V} \\
-\sin\theta_{V} & \cos\theta_{V}
\end{array}    \right )
\left (  \begin{array}{c} \phi \\
     \omega
\end{array}
\right ) ,
\end{eqnarray}
and the mixing angle has been fixed to be $\theta_V=38.62^\circ$ following Refs.~\cite{Wang:2023njt,Zhang:2025ijd}.

\subsection{\texorpdfstring%
{TFF for $J/\psi\to \pi^0\gamma^*$}%
{J/psi to pi0 gamma* transition form factor}}
\begin{figure}[!htb]
\centering
\includegraphics[width=1\linewidth]{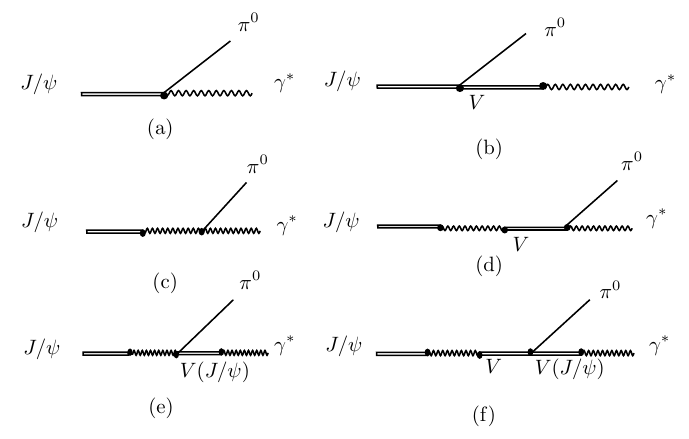}
\caption{Feynman diagrams of $J/\psi\to \pi^0\gamma^*$. The LO results are shown in the first row, and the electromagnetic correction and charmonium contribution are listed in the other rows. The double line represents vector resonance.}
\label{Fig:Feynman1}
\end{figure}

The decay amplitude of $J/\psi\to\pi^0(p)\gamma^*(q)$ is given as
\begin{eqnarray}
    \mathcal{M} _{\psi\pi^0\gamma^*} &=&e\varepsilon _{\mu \nu \rho \sigma } \epsilon ^{\mu}_{\psi} \epsilon ^{\nu}_{\gamma^*} k^{\rho }q^{\sigma } G_{\psi\pi^0\gamma^*}(q^2), \nonumber
\end{eqnarray}
where $G_{\psi\pi^0\gamma^*}(q^2)$ is the TFF, and it can be written as
\begin{eqnarray}
    G_{\psi\pi^0\gamma^*}(q^2)=G^{\mathrm{LO}}_{\psi\pi^0\gamma^*}(q^2)+G^{\mathrm{EM}}_{\psi\pi^0\gamma^*}(q^2)+G^{c\bar{c}}_{\psi\pi^0\gamma^*}(q^2). \nonumber\\ \label{Eq:J;TFF}
\end{eqnarray}
The superscripts ``LO, EM, and $c\bar{c}$" represent the LO strong interaction, EM transition, and charmonium contributions, respectively.
The Feynman diagrams of $J/\psi\to \pi^0\gamma^*$ are shown in Fig.~\ref{Fig:Feynman1}.
The form factors are calculated out within RChT, where one has
\begin{eqnarray}  \label{Eq:J;TFF2}
    G_{\psi\pi^0\gamma\!^*}^{\rm LO}\!(q^2\!)&=&
    \frac{-4}{FM_\psi^2}\! (g^{\psi}_{1}\! M_\psi^2\!\!+\!\!4g^{\psi}_{2} m_\pi^2) \nonumber\\
   &+&\!\!\frac{2\sqrt{2} }{FM_\psi\!}\! (h^{\psi}_1 \!M_\psi^2\!\!+\!\!4h^{\psi}_2\!m_\pi^2\!)\!
   \Big[F\!_\rho(q^2)\cos\!{\delta}\mathrm{BW}(\rho,q^2) \nonumber\\
   &-&\!F_\omega(q^2)\sin{\delta_\omega}(q^2) \mathrm{BW} (\omega,q^2)\Big].\nonumber\\
    G_{\psi\pi^0\gamma^*}^{\mathrm{EM}}(q^2)&=&\frac{8\sqrt{2}\pi\alpha f_\psi}{3M_\psi}\mathcal{F}_{\pi^0\gamma^*\gamma^*}(M_\psi^2,q^2).\nonumber\\
    G_{\psi\pi^0\gamma^*}^{c\bar{c}}(q^2)&=&G_{\psi\pi^0\gamma^*}^{c\bar{c}-\mathrm{IB}}(q^2)+G_{\psi\pi^0\gamma^*}^{c\bar{c}-\mathrm{EM}}(q^2)\,.
\end{eqnarray}
The form factors $F\!_{V}(q^2)$ are given as
\begin{eqnarray}
F\!_\rho(q^2)\!&=&\!\frac{F_V}{9} \bigg[9 \cos\!\delta \big(1\!+\! \frac{8\sqrt{2}\alpha _{V} }{M_V^2}  m_{\pi }^2\big)
\!+\!\sqrt{3} \sin\!\delta_\rho(q^2) \nonumber\\
&&\big(3  \sin\!\theta_V \!-\! \frac{16\sqrt{2}\alpha _{V} }{M_V^2}  m_K^2 (\!\sqrt{2} \cos\! \theta_V\!-\!2 \sin\!\theta_V \!) \nonumber\\
\!&+&\! \frac{8\sqrt{2}\alpha_{V} }{M_V^2}  m_{\pi }^2 (2 \sqrt{2} \cos\!\theta_V\!-\!\sin\!\theta_V \!)\big)\!\bigg],\nonumber\\
F\!_\omega(q^2)\!&=&\!\frac{F_V}{9} \bigg[\sqrt{3} \cos\!\delta \Big(\!3  \sin\!\theta_V \!-\! \frac{16\sqrt{2}\alpha _{V} }{M_V^2}  m_K^2 \big(\!\sqrt{2} \cos\! \theta_V\!\nonumber\\
&&-\!2 \sin\!\theta_V \!\big) \!+\! \frac{8\sqrt{2}\alpha _{V} }{M_V^2}  m_{\pi }^2 \big(\!2 \sqrt{2} \cos\!\theta_V\!-\!\sin\!\theta_V \!\big)\!\Big) \nonumber\\
\!& &\! -9 \sin\!\delta_\omega(q^2) \big(1\!+\! \frac{8\sqrt{2}\alpha _{V} }{M_V^2}  m_{\pi }^2\big)\bigg]\nonumber\\
F\!_\phi(q^2)\!&=&\! \frac{F_V}{3\sqrt{3}} \bigg[3 \cos\!\theta_V \!+\! \frac{16\sqrt{2}\alpha_V}{M_V^2} m_K^2 \big(\!\sqrt{2} \sin\!\theta_V\!+\!2 \cos\!\theta_V \!\big) \!\nonumber\\
\!& &\!- \! \frac{8\sqrt{2}\alpha_V}{M_V^2} m_{\pi}^2 \big(\!2\sqrt{2} \sin\!\theta_V \!+\! \cos\!\theta_V \!\big) \bigg],
\end{eqnarray}
with $F_V=\sqrt{3}F$ \cite{Roig:2014}.


Our analysis addresses the higher energy region, where the heavier vector resonance multiplets, $V'$ and $V''$, must be accounted for. Therefore, we apply the extended BW propagators following Refs.~\cite{Wang:2023njt,Zhang:2025ijd},
\begin{equation}
\mathrm{BW}(V,x)= \frac{1}{\Delta _{V}(x) } \longrightarrow \frac{1}{\Delta _{V}(x) } +\frac{\beta _{X}'}{\Delta _{V'}(x) } +\frac{\beta _{X} ''}{\Delta _{V''}(x) } , \label{Eq:BW}
\end{equation}
where one has $X=\psi\pi^0\gamma^*$ for $G^{\mathrm{LO}}_{\psi\pi^0\gamma^*}(q^2)$.
The off-shell width of the resonance $\rho$ is based on the formulas of Refs.~\cite{GomezDumm:2000fz,Dai:2013joa,Wang:2023njt},
\begin{eqnarray}
 \Gamma\!_\rho\!\!\left(q^2\right)\!&=&\!\frac{M_\rho q^2}{96 \pi F^2}\!\!\left[\sigma_\pi^3\left(q^2\right)\! \theta\left(\!q^2\!\!-\!\!4 m_\pi^2\right) \Lambda[q^2,M_\rho,m_\pi,m_\pi]\right.\nonumber\\
 &&\left.
  \!+\!\!\frac{1}{2} \sigma_K^3\left(q^2\right) \theta\left(q^2\!\!\!-\!\!4 m_K^2\right)\Lambda[q^2,M_\rho,m_K,m_K]\!\right]\!,\nonumber
\end{eqnarray}
Where $\theta(x)$ is the step function, and $\sigma_P(x)=\sqrt{1-4m_P^2/x}$ is the phase space factor.
A Blatt-Weisskopf form factor~\cite{Blatt:1952ije,ParticleDataGroup:2024cfk} is introduced to enforce the correct asymptotic behavior of the BW propagator,
\begin{eqnarray}
 \Lambda[x,M,m_a,m_b]&=&\frac{1+\big(q_{ab}(M^2,m_a^2,m_b^2)\cdot R)^2}{1+\big(q_{ab}(x,m_a^2,m_b^2)\cdot R)^2}\,,
 \end{eqnarray}
where one has
\begin{equation}
q_{ab}(q^2,m_a^2,m_b^2)=\frac{1}{2}\sqrt{\frac{\lambda(q^2,m_a^2,m_b^2)}{q^2}}\,, \nonumber
\end{equation}
with $\lambda(x,y,z)=x^2+y^2+z^2-2xy-2xz-2yz$. For the case of two equal masses, the expression simplifies to $q_{aa}(q^2,m_a^2,m_a^2)=\sqrt{q^2-4m_a^2}/2$. $R$ is the hadronic scale and one can set $R=1.5~\mathrm{GeV}^{-1}$ for all vector resonances as Ref.~\cite{Belle:2004drb}. In practice, this value gives a good description of the data in our analysis.
Given the narrow widths of the $\omega$ and $\phi$ resonances, we treat their widths as constants in our TFFs, multiplying a step function. One has
\begin{eqnarray}
    \Gamma_{\omega/\phi}(q^2)&=&\Gamma_{\omega/\phi}\,\theta(q^2-{\rm th}_{\omega/\phi}),
\end{eqnarray}
where one has ${\rm th}_{\omega}=9m_\pi^2$ and  ${\rm th}_{\phi}=4m_K^2$.
Here and after, the $\Gamma_{V,V',V''}$ are the physical decay widths (constants) of the vectors.
The off-shell widths of heavier vector resonances, $V'$ and $V''$, are parameterized by momentum dependent forms \cite{Dai:2013joa,Wang:2023njt},
\begin{eqnarray}
    \Gamma_{V'('')}(q^2)&=&\Gamma_{V'('')}\frac{q^3_{12}(q^2,m_1^2,m_2^2)}{q^3_{12}(M_{V'('')}^2,m_1^2,m_2^2)}\theta(q^2-\mathrm{th}_{V'('')}) \nonumber \\
    &\times&\Lambda[q^2,M_{V'('')},m_1,m_2],
\end{eqnarray}
where the threshold is given as $\mathrm{th}_{V'('')}=(m_1+m_2)^2$. The subscripts of the mass are $1,2=\pi\pi$, $\pi\rho$, and $K \bar{K}$ for $\rho'('')$, $\omega'('')$, and $\phi'('')$, respectively.


As shown in Figs.~\ref{Fig:Feynman1} (c-f), the $G_{\psi\pi^0\gamma^*}^{\mathrm{EM}}(q^2)$ involves the time-like doubly-virtual pion TFF, $\mathcal{F}_{\pi^0\gamma^*\gamma^*}(q_1^2,q^2)$. The pion TFF can be defined through the  $\pi^0\to\gamma^*(q_1)\gamma^*(q_2)$ decay amplitude,
\begin{eqnarray}
\mathcal{M} _{\pi^0\gamma ^{*} \gamma ^{*}  } &=&ie^2\varepsilon ^{\mu \nu \rho \sigma } q_{1\mu } q_{2\nu }\epsilon _{1\rho } \epsilon _{2\sigma }  \mathcal{F} _{\pi^0 \gamma^* \gamma^{*}  }(q_1^2,q_2^2)\,.\nonumber
\end{eqnarray}
The detailed expressions of the pion TFF can be found in Ref.~\cite{Zhang:2025ijd}. For the one used in our work, $\mathcal{F}_{\pi^0\gamma^*\gamma^*}(M_\psi^2,q^2)$, $M_\psi^2$ lies outside the working range where our previous pion TFF is valid, $q_{1,2}^2<2.3^2~\mathrm{GeV}^2$. Therefore, an extension of the pion TFF is required. In line with the asymptotic behavior predicted by pQCD~\cite{Hoferichter:2020lap}, $\mathcal{F}^{\mathrm{pQCD}}_{\pi^0\gamma^*\gamma^*}(q_1^2,q_2^2)\propto  \frac{1}{q_1^2+q_2^2}$, we can apply the following scheme,
\begin{eqnarray}\label{Eq:TFF;con}
  \mathcal{F}_{\pi^0\gamma^*\gamma^*}(M_\psi^2,q^2)\!=\!\!  \begin{cases}
\mathcal{F}^{\mathrm{RChT}}_{\pi^0\gamma^*\gamma^*}(s_0,q^2) \frac{s_0+q^2}{M_\psi^2+q^2} &\!\!\!\!\!\! \text{ if } q^2\le s_0 \\
\mathcal{F}^{\mathrm{RChT}}_{\pi^0\gamma^*\gamma^*}(s_0,s_0)\frac{2s_0}{M_\psi^2+q^2} &\!\!\!\!\!\! \text{ if } q^2>  s_0.
\end{cases}\nonumber\\
\end{eqnarray}
Given that our previous pion TFF~\cite{Zhang:2025ijd} successfully describes data up to $q_{1,2}^2 \approx 2.3^2$ GeV$^2$, we set $s_0 = 2.3^2~\mathrm{GeV}^2$ as the cutoff. This value indeed results in a good fit to the $\psi\pi^0\gamma^*$ TFF.

According to Eq.~(\ref{Eq:J;TFF2}), the charmonium contributions split into electromagnetic transitions and strong interactions, with the latter accounting for isospin-breaking (IB) effects. To estimate the strong-interaction part (IB), we adopt a simple monopole ansatz~\cite{Fu:2011yy,Kubis:2014gka}.   One has
\begin{eqnarray}
        G^{c\bar{c}-\mathrm{EM}}_{\psi\pi^0\gamma^*}(q^2)&=&\frac{32\pi\alpha f_\psi^2 q^2 \mathrm{BW} (\psi,q^2)}{9M_\psi^2}G^{\mathrm{LO}}_{\psi\pi^0\gamma^*}(q^2)\,,\nonumber\\
    G_{\psi\pi^0\gamma^*}^{c\bar{c}-\mathrm{IB}}(q^2)&=&\frac{G_{\psi\pi^0\gamma^*}^{c\bar{c}-\mathrm{IB}}(0)}{1-q^2/\Lambda^2}\,.
\end{eqnarray}
Here, the superscript \lq$c\bar{c}$-IB' represents isospin-breaking effects in the charmonium contribution. In practice, it is found that using $G_{\psi\pi^0\gamma^*}^{c\bar{c}-\mathrm{IB}}(0) = 0.95 \times 10^{-4}~\mathrm{GeV}^{-1}$ and $\Lambda = (M_\psi + M_{\psi'})/2$ yields a high quality description to the data, as discussed in the following section.

The branching ratio of $J/\psi\to\pi^0\gamma$ is given as
\begin{equation}
    \mathrm{Br}[{\psi\to\pi^0\gamma}]=\frac{\alpha(M_{\psi}^2-m_{\pi^0}^2)^3}{24M_{\psi}^3\Gamma_\psi}|G_{\psi\pi^0\gamma^*}(0)|^2,
\end{equation}
and the differential branching ratio of $J/\psi\to\pi^0l^+l^-$ is
\begin{eqnarray}
    \frac{\mathrm{d}\mathrm{Br}[{\psi\to\pi^0l^+l^-}]}{\mathrm{d}\sqrt{s}}&=&\frac{\alpha^2(s+2m_l^2)}{36\pi M_\psi^3 s^2\Gamma_\psi}\sqrt{s-4m_l^2}\nonumber\\&\times&\lambda^{3/2}(M_\psi^2,m_\pi^2,s)|G_{\psi\pi^0\gamma^*}(s)|^2.\nonumber\\
\end{eqnarray}
Here, $l$ denotes either the electron or the muon, and $s=(p_{l^+}+p_{l^-})^2$ is the Mandelstam variable.
The LO contribution dominates $\Gamma_{J/\psi\to\pi^0\gamma}$ and $\Gamma_{J/\psi\to\pi^0e^+e^-}$, although the EM transition also play a crucial role in the $\rho-\omega$ and $\phi$ energy regions. In particular, the LO contribution does not include the $\phi$ resonance due to the OZI rules \cite{Lipkin:1984sw}. Thus, the structure around the $\phi$ energy region should originate from the EM transition. See discussions below.

\subsection{\texorpdfstring%
{$J/\psi\to \pi^0 V$ decay widths}%
{J/psi to pi0 V decay widths}}
\begin{figure}[htbp]
    \includegraphics[width=0.48\textwidth]{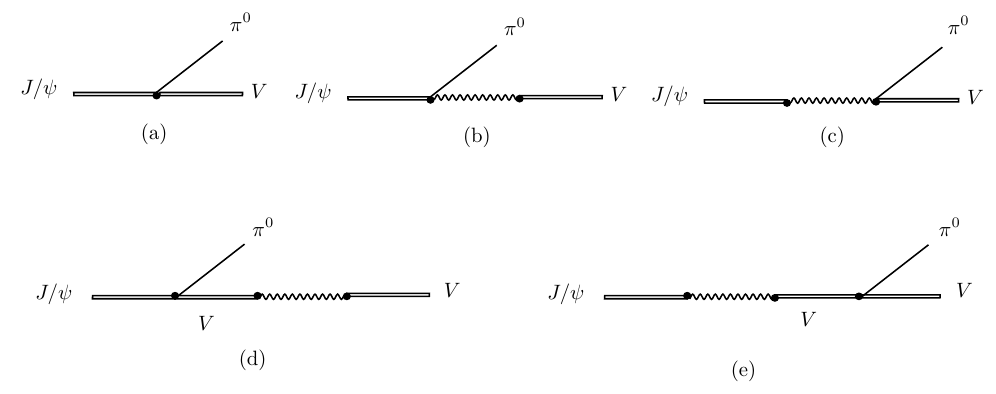}
    \caption{Feynman diagrams of $J/\psi\to \pi^0V$. The LO result is from (a), and the electromagnetic transition are from (b-e). }
    \label{Fig:Feynman2}
\end{figure}

The decay widths of $J/\psi\to \pi^0(p) V(q)$ can be included to refine our analysis.
This is also essential for clarifying the dominant dynamics in each decay channel, $J/\psi\to\pi^0 \rho^0$, $\pi^0 \omega$, and $\pi^0 \phi$.
The corresponding Feynman diagrams are shown in Fig.~\ref{Fig:Feynman2}.
We note that in Fig.~\ref{Fig:Feynman2} (d), the internal vector line cannot be the same as the final-state one; otherwise it would be absorbed into the re-definition of the field through wave-function renormalization.
The decay amplitude is given as
\begin{equation}
    \mathcal{M} _{\psi V\pi^0} =e\varepsilon _{\mu \nu \rho \sigma } \epsilon ^{\mu}_{\psi} \epsilon ^{\nu}_{\gamma^*} k^{\rho }q^{\sigma } G_{\psi V\pi^0}\,, \nonumber
\end{equation}
where the form factor $G_{\psi V\pi^0}$ is given as
\begin{equation}
    G_{\psi V\pi^0}=G_{\psi V\pi^0}^{\mathrm{LO}}+G_{\psi V\pi^0}^{\mathrm{EM}}.
\end{equation}
The LO form factors $G_{\psi V\pi^0}^{\mathrm{LO}}$ for each channel ($V=\rho^0,~\omega,~\phi$) are given as
\begin{eqnarray}
G_{\psi \rho^0\pi^0}^{\mathrm{LO}}&=&\frac{2\sqrt{2}\cos{\delta}}{FM_\rho M_\psi}(h_{1}^{\psi}M_\psi^2+4h_2^{\psi}m_\pi^2)\,,\nonumber\\
G_{\psi \omega\pi^0}^{\mathrm{LO}}&=&-\frac{2\sqrt{2}\sin{\delta_{\omega}(M_\omega^2)}}{FM_\omega M_\psi}(h_{1}^{\psi}M_\psi^2+4h_2^{\psi}m_\pi^2)\,,\nonumber\\
G_{\psi \phi\pi^0}^{\mathrm{LO}}&=&0\,,
\end{eqnarray}
and the EM form factor $G_{\psi V\pi^0}^{\mathrm{EM}}$ is given as
\begin{eqnarray}
    G_{\psi V\pi^0}^{\mathrm{EM}}&=&\frac{-8\pi\alpha F_V(M_V^2)}{FM_V }\bar{G}_{\psi\pi^0\gamma^*}^{\mathrm{LO}-V}(M_V^2)\nonumber\\&&+\frac{8\sqrt{2}\pi\alpha f_\psi}{3M_\psi}F_{V\pi^0\gamma^*}(M_\psi^2)\,.
\end{eqnarray}
Here, $\bar{G}_{\psi\pi^0\gamma^*}^{\mathrm{LO}-V}(q^2)$ for $V=\rho^0,~\omega,~\phi$ are given as
\begin{eqnarray}
    \bar{G}_{\psi\pi^0\gamma^*}^{\mathrm{LO}\!-\!\rho^0}(q^2)\!&=&\!
    \frac{-4}{F\!M\!_\psi\!\!^2}\! (g^{\psi}_{1}\! M\!_\psi\!\!^2\!\!+\!\!4g\!^{\psi}_{2} \!m_\pi^2\!)
   \!-\!\frac{2\sqrt{2} }{F\!M_\psi\!}\! (h^{\psi}_1 \!M_\psi^2\!\!+\!\!4h^{\psi}_2\!m_\pi^2\!)\nonumber\\
   &\times&\!F_\omega(q^2)\!\sin{\!\delta_\omega}(q^2) \mathrm{BW} (\omega,q^2)\,, \nonumber\\[2mm]
    \bar{G}_{\psi\pi^0\gamma^*}^{\mathrm{LO}-\omega}(q^2)\!&=&\!
    \frac{-4}{F\!M\!_\psi\!\!^2}\! (g^{\psi}_{1}\! M\!_\psi\!\!^2\!\!+\!\!4g\!^{\psi}_{2} \!m_\pi^2\!) \!+\!\frac{2\sqrt{2} }{F\!M_\psi\!}\! (h^{\psi}_1 \!M_\psi^2\!\!+\!\!4h^{\psi}_2\!m_\pi^2\!)\!\nonumber\\
   &\times&F\!_\rho(q^2)\cos\!{\delta}\mathrm{BW} (\rho,q^2)\,, \nonumber\\[2mm]
    \bar{G}_{\psi\pi^0\gamma^*}^{\mathrm{LO}-\phi}(q^2)&=&G_{\psi\pi^0\gamma^*}^{\mathrm{LO}}(q^2).
\end{eqnarray}
The form factor $F_{V\pi^0\gamma^*}(q^2)$ can be calculated from the $V\to \gamma^*(q)\pi^0$ amplitude,
\begin{equation}
    \mathcal{M}_{V\to \pi^0\gamma^*}=-e\varepsilon _{\mu \nu \rho \sigma } \epsilon ^{\mu}_{V} \epsilon ^{\nu}_{\gamma^*} k^{\rho }q^{\sigma }F_{V\pi^0\gamma^*}(q^2)\,, \nonumber
\end{equation}
and the explicit forms for $V=\rho^0,~\omega,~\phi$ are given as
\begin{eqnarray}
    F\!_{\!\rho\!^0\pi\!^0\gamma^*}\!(q^2\!)\!&=&\!\!
    \frac{-2}{3FM_V M_\rho}\!\!\big[\!\sqrt{2}\!\cos{\!\delta}\!\!+\!\!\sqrt{3}(2\!\cos{\!\theta_V}\!\!+\!\!\sqrt{2}\!\sin{\!\theta_V})\nonumber\\&&\sin\!{\delta\!_\rho}\!(M\!_\rho\!\!^2) \big] (\tilde{c}_{125}q^2+\tilde{c}_{1235}m_\pi^2-\tilde{c}_{1256}M\!_\rho\!\!^2)\nonumber\\
    \!&+&\!\!\frac{2}{3F\! M\!_\rho}\!\!\big[ 2 \mathrm{BW}\!(\phi,\!q^2\!)\!\cos{\!\delta}F\!\!_\phi(q^2\!)\!\big(\!\cos{\!\theta\!_V}\!\!-\!\!\sqrt{2}\!\sin{\!\theta\!_V} \!\big) \nonumber\\
    &+&\!\big(\! \sqrt{2}\!\cos{\!\theta\!_V}\!+\!\sin{\!\theta\!_V}\! \big)\big( 2\mathrm{BW}\!(\!\rho,q^2\!)\cos{\!\delta}F\!_\rho(\!q^2)\nonumber\\
    &&\!\!\!\! (\sin{\!\delta\!_\rho}(q^2)\!+\!\sin{\!\delta\!_\rho}(M_\rho^2)) \big)
    \!\!\!+\!\mathrm{BW}(\!\omega,q^2)F\!_\omega\!(\!q^2)\nonumber\\
    &&\!\!\!\!\!\big( 1+\cos{2\delta}-2\sin{\delta_\rho}(M_\rho^2)\sin{\delta\!_\omega}(q^2) \big)\big]\nonumber\\
    &\times& \big( \tilde{d}_{123}m_\pi^2+\tilde{d}_3 (q^2+M\!_\rho\!\!^2) \big). \nonumber\\[2mm]
F\!_{\omega\pi^0\gamma^*}(q^2)\!\!&=&\!\!\!
    \!\frac{-2}{3FM_V M_\omega}\big[\sqrt{3}\cos{\delta}(2\cos{\!\theta\!_V}\!+\!\sqrt{2}\sin{\!\theta\!_V})\nonumber\\&-&\sin{\!\delta\!_\omega}\!(\!M_\omega^2) \big](\tilde{c}_{125}q^2+\tilde{c}_{1235}m_\pi^2-\tilde{c}_{1256}M_\rho^2)\nonumber\\
    &+&\!\frac{2}{\sqrt{3}F M_\omega}\big[\!2 \mathrm{BW}(\phi,q^2)\sin{\delta_\omega}(M_\omega^2) F_\phi(q^2)\nonumber\\
    &&\big(\!\sqrt{2}\!\sin{\!\theta\!_V}\!\!-\!\cos{\!\theta\!_V}\! \big)\!
    \!-\!\!\big(\! 2\mathrm{BW}\!(\!\omega,q^2\!)\!\cos{\delta}F_\omega(q^2)\nonumber\\
    &&(\!\sin{\!\delta\!_\omega}\!(\!q^2)\!+\!\sin{\!\delta\!_\omega}\!(\!M_\omega^2)) \big)\!\!\!
    +\mathrm{BW}(\rho,q^2)F_\rho(q^2)\nonumber\\&&\big( 1+\cos{2\delta}-2\sin{\delta_\rho}(q^2)\sin{\delta_\omega}(M_\omega^2) \big)\nonumber\\&&
    (\sqrt{2}\!\cos{\theta_V}\!+\!\sin{\theta_V})\big]
    \big( \!\tilde{d}_{123}m_\pi^2\!\!+\!\!\tilde{d}_3\! (\!q^2\!\!\!+\!M_\omega^2)\!\big).\nonumber\\[2mm]
    F\!_{\phi\pi^0\gamma^*}(q^2)\!\!&=&\!
    \frac{-2}{\sqrt{3}F\!M\!_V M\!_\phi}(\sqrt{2}\!\cos{\!\theta\!_V}-2\sin{\!\theta\!_V}) (\tilde{c}_{125}q^2\nonumber\\&+&\!\!\tilde{c}_{1235}m_\pi^2\!-\!\tilde{c}_{1256}M_\phi^2)\!+\!\frac{4}{\sqrt{3}F M_\phi}
    \big[\mathrm{BW}(\rho,q^2)\nonumber\\
    &&\!\!\!\!\!\!\cos{\delta}F_\rho(q^2)-\mathrm{BW}(\omega,q^2)\sin{\delta_\omega}(q^2)F_\omega(q^2)\big]
    \nonumber\\
    &&\!\!\!\!\!\!(\cos{\!\theta\!_V}\!\!-\!\!\sqrt{2}\sin{\!\theta\!_V})\big( \tilde{d}_{123}m_\pi^2+\tilde{d}_3 (q^2+M_\phi^2) \big).\nonumber\\
\end{eqnarray}
The definitions of the parameters $\tilde{c}$ and $\tilde{d}$ are given in Refs. \cite{Dai:2013joa, Guevara:2018rhj}.

The branching ratio of $J/\psi\to V\pi^0$ is given as
\begin{equation}
    \mathrm{Br}[{J/\psi\to \pi^0 V}]=\frac{\lambda^{3/2}(M_\psi^2,M_V^2,m_\pi^2)}{96\pi M_{\psi}^3\Gamma_\psi}|G_{\psi V\pi^0}|^2.
\end{equation}

\section{Fit results and discussions}\label{sec3}
In this work, the parameters from the pion TFF are taken from our previous work \cite{Zhang:2025ijd}, including the masses and widths of the vector resonances.  Considering that the $V''$ contribution should be smaller than that of $V'$ \cite{Dai:2013joa,Qin:2020udp,Wang:2023njt,Qin:2024ulb} and given the limited data above 1.5 GeV~\cite{BESIII:2025xjh}, we set $\beta_{\psi\pi^0\gamma^*}''=0$. In practice, the fit quality remains high without this parameter, as discussed later. The only remaining unkonwn parameters are $g^\psi_{12}=\frac{(g^\psi_1M_\psi^2+4g^\psi_2m_\pi^2)}{M_\psi^2}$, $h^\psi_{12}=\frac{(h^\psi_1M_\psi^2+4h^\psi_2m_\pi^2)}{M_\psi^2}$, $\beta_{\psi\pi^0\gamma^*}'$. The fit is preformed using MINUIT \cite{James:1975dr}, and the resulting parameters are listed in Table \ref{tab:para}.
\begin{table}[hptb!]
\centering
{\footnotesize
\renewcommand{\arraystretch}{2}
\newcommand{\tabincell}[2]{\begin{tabular}{@{}#1@{}}#2\end{tabular}}
\begin{tabular}{|c|c|c|c|}
\colrule
 $g^\psi_{12}~(10^{-6})$ & $h^\psi_{12}~(10^{-6})$ &$\beta_{\psi\pi^0\gamma*}'$ &$\chi^2_{\mathrm{d.o.f.} }$ \\
\colrule
$-2.6\pm 3.1$ & $-29.5\pm0.7$  & $0.9\pm0.7$ &0.91  \\
\colrule
\end{tabular}
\caption{Fitting parameters. The uncertainties of the parameters are taken from MINUIT \cite{James:1975dr}.}
\label{tab:para}
}
\end{table}

\begin{figure}[htbp]
    \includegraphics[width=0.48\textwidth,height=0.38\textheight]{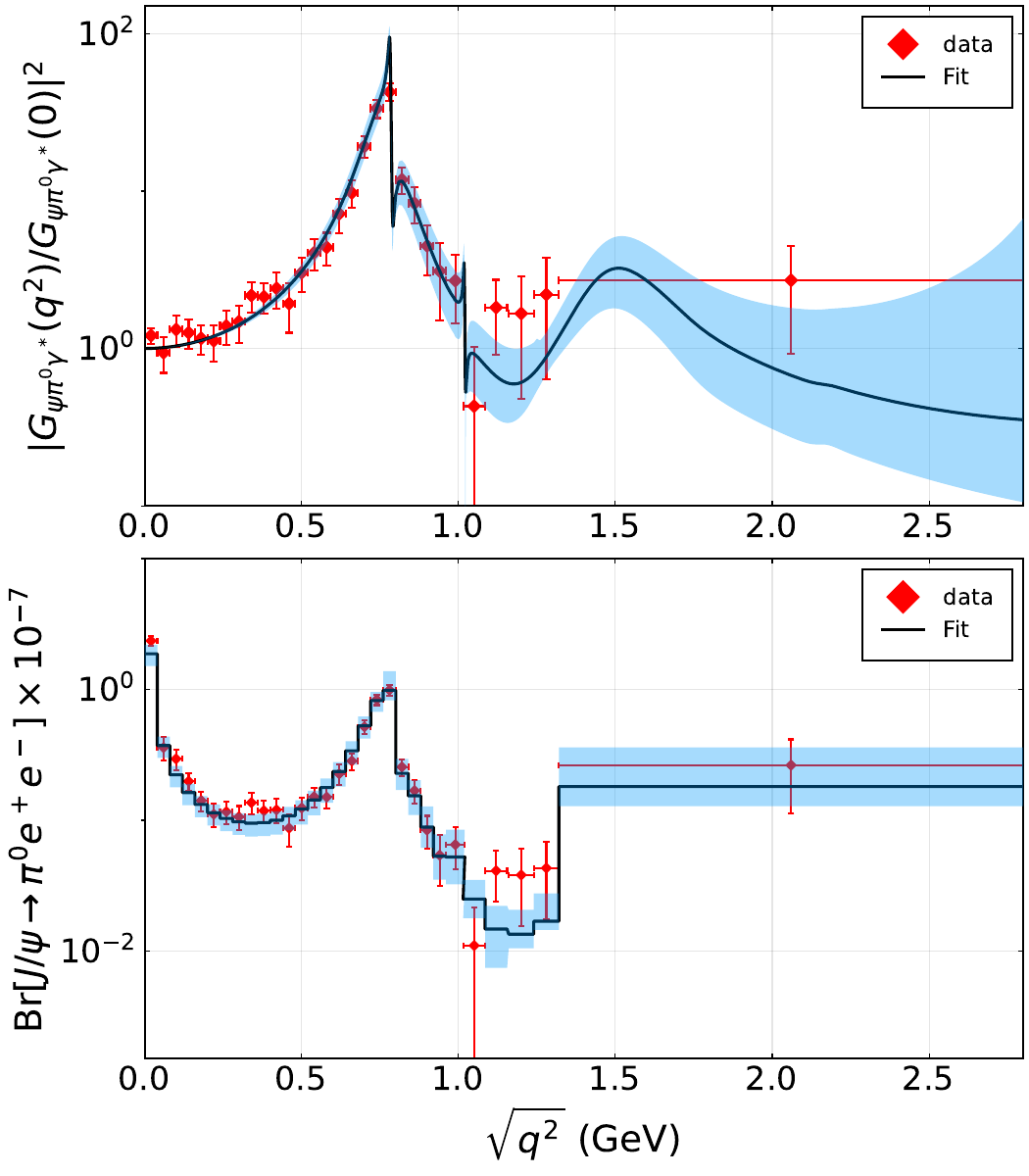}
    \caption{Normalized TFF and $e^+e^-$  invariant mass spectrum for $J/\psi\to\pi^0 e^+e^-$. The data is taken from Ref.~\cite{BESIII:2025xjh}.}
    \label{Fig:fit}
\end{figure}

The fitting results for the $e^+e^-$ invariant mass spectrum of $J/\psi\to\pi^0 e^+e^-$ are shown in Fig.~\ref{Fig:fit}.
The errors of the physical observables are estimated using the bootstrap method \cite{Efron:1979bxm}, where random variations are applied to the data points based on a normal distribution of their central values and uncertainties.
We also account for the uncertainties associated with the charmonium contribution. They are estimated by varying $G_{\psi\pi^0\gamma^*}^{c\bar{c}-\mathrm{IB}}(0)$ within the range $[0.3, 1.6] \times 10^{-4}~\mathrm{GeV}^{-1}$ as suggested in Ref.~\cite{Kubis:2014gka}, and $\Lambda$ over the interval $[M_\psi, M_{\psi'}]$.

As can be found in Fig.~\ref{Fig:fit}, our model provides a good description of the latest data~\cite{BESIII:2025xjh} with only three free parameters, yielding $\chi^2_{\mathrm{d.o.f.} }=0.91$. See Table \ref{tab:para}.
Especially, in the low energy region, $E_{\gamma^*}=\sqrt{q^2}\leq 1.2~\mathrm{GeV}$, only two parameter $g_{12}^\psi$ and $h_{12}^\psi$ have significant effects, because the contributions from  $\beta_{\psi\pi^0\gamma^*}'$ predominantly influence the higher energy region.
Note that $g_{12}^\psi$ and $h_{12}^\psi$ are strongly constrained by the decay width/branching ratio of $J/\psi\to\pi^0\gamma$, leaving limited freedom for their values.
Hence, the high quality fit indicates that the continued doubly-virtual pion TFF $\mathcal{F}^{\mathrm{pQCD}}_{\pi^0\gamma^*\gamma^*}(q_1^2,q_2^2)$,  as given by Eq.~(\ref{Eq:TFF;con}), works rather well even at the large virtuality, $q_1^2=M_\psi^2$.
Near the $e^+e^-$ threshold, our model describes the data of branching ratio fairly well. See the first data point in the second graph of Fig.~\ref{Fig:fit}. The reason is that $G_{\psi\pi^0\gamma^*}(q^2)$ near threshold is dominanted by $\Gamma_{\psi\to\pi^0\gamma}$, There is a tension between this decay width and the first bin data, as the former corresponds to a smaller value of $|G_{\psi\pi^0\gamma^*}(0)|$. If the first data point is excluded, the $\chi^2_{\mathrm{d.o.f.} }$ decreases to $0.71$.
Our fit is also not perfect in the energy region around 1.2~GeV, but the deviations remain within the experimental uncertainties. There are distinct structures around the lightest vector resonances, $\rho$, $\omega$, and $\phi$. It is important to note, however, that the dynamics producing these resonances differ: some originate from the LO strong interactions, while others arise from electromagnetic transitions. This will be discussed in detail below.

Our predictions of branching ratios are shown in Table~\ref{tab:br}.
\begin{table}[htbp]
{\footnotesize
\renewcommand{\arraystretch}{2}
\newcommand{\tabincell}[2]{\begin{tabular}{@{}#1@{}}#2\end{tabular}}
\begin{ruledtabular}
\begin{tabular}{cccc}
Br & This work & Others & Data \\
\colrule
\tabincell{c}{$J/\psi\to\pi^0\gamma$\\$(10^{-5})$} & $3.42\!\pm\! 0.08$ & \tabincell{c}{ $3.03\!\pm\!0.86$\cite{Chen:2014yta} \\$3.41\!\pm\!0.16$\cite{Yan:2023nqz}  } & $3.39\pm0.08$\cite{ParticleDataGroup:2024cfk} \\
\colrule
\tabincell{c}{$J/\psi\to\pi^0e^+e^-$\\$(10^{-7})$} & $7.70\!\pm\!0.57$  &  \tabincell{c}{ $11.91\!\pm\!1.38 $\cite{Chen:2014yta} \\$12.94\!\pm\!0.44$\cite{Yan:2023nqz} \\ $(5.5...6.4)$\cite{Kubis:2014gka}\\$3.89^{+0.37}_{-0.33}$\cite{Fu:2011yy} \\$7.03^{+0.62}_{-0.60}$\cite{Cao:2025ncx}  } &  \tabincell{c}{ $8.06\!\pm\!0.49$\cite{BESIII:2025xjh} \\$7.6\!\pm\!1.4$\cite{ParticleDataGroup:2024cfk} }\\
\colrule
\tabincell{c}{$J/\psi\to\pi^0 \mu^+\mu^-$\\$(10^{-7})$}& $4.80 \!\pm\!0.38$ &  \tabincell{c}{$2.80\!\pm\!0.32$\cite{Chen:2014yta} \\$3.04\!\pm\!0.10$\cite{Yan:2023nqz} \\ $(2.7...3.3)$\cite{Kubis:2014gka}\\$1.01^{+0.10}_{-0.09}$~\cite{Fu:2011yy} \\$4.15^{+0.58}_{-0.50}$\cite{Cao:2025ncx}   } & - \\
\end{tabular}
\end{ruledtabular}
\caption{Prediction of branching ratios from our model. }
\label{tab:br}
}
\end{table}
As can be found, our results of ${\rm Br}[{J/\psi\to\pi^0\gamma}]$ and ${\rm Br}[{J/\psi\to\pi^0 e^+e^-}]$ agree well with the experimental values \cite{ParticleDataGroup:2024cfk,BESIII:2025xjh}.  The first branching ratio is included as a contraint in our fit, so the close agreement is expected. The second is a prediction of our model, and its consistency with data supports the reliability of our model. The third branching ratio is also a prediction and can be tested in future experiments.

\begin{figure}[htbp!]
    \includegraphics[width=1\linewidth]{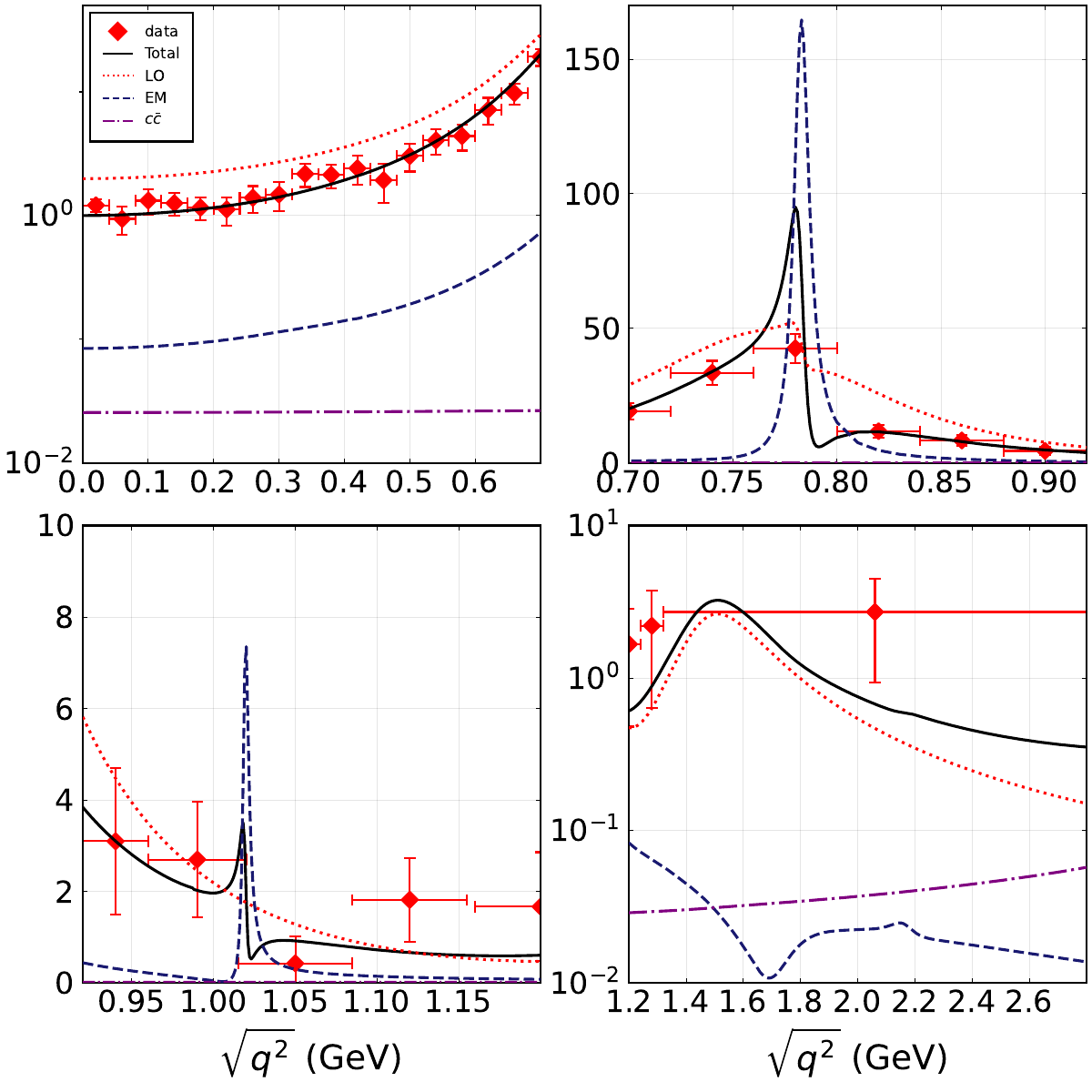}
    \caption{The \lq LO', \lq EM', and \lq $c\bar{c}$' contributions to the TFF of $\psi\pi^0\gamma^*$.  }
    \label{Fig:ind}
\end{figure}

The individual contributions to the $e^+e^-$ invariant mass spectrum arising from each term of Eq.~(\ref{Eq:J;TFF}) are presented in Fig.~\ref{Fig:ind}.
At energies below the $\rho$-$\omega$ mass region, as shown in the upper-left graph of Fig.~\ref{Fig:ind}, the dynamics are dominated by the LO contribution. In contrast, the charmonium contribution is nearly flat and negligible compared to other contributions.
As the energy increases, the EM transition contribution grows significantly and becomes comparable to the LO contribution. See the upper-right graph of Fig.~\ref{Fig:ind}. In the resonance ($\rho$-$\omega$) region, the EM contribution (blue dashed line) even exceeds  the LO one (red dotted line). Moreover, the distinct structure in this region can not be produced by a single BW resonance, indicating that the $\omega$ contribution should not be ignored.
According to isospin conservation, the subsequent decay, $J/\psi\to\pi^0\rho^0\to\pi^0\gamma^*$ dominates the LO contribution, and the $J/\psi\to\pi^0\omega\to\pi^0\gamma^*$ process is suppressed as it arises from isospin symmetry breaking.
Therefore, the $\omega$ contribution originates not only from the IB effect, but more importantly from the EM transition represented by  Figs.~\ref{Fig:Feynman1} (c-f).
This intepretation is consistent with the observed structure around the $\phi$ resonance region, as shown in the lower-left graph of Fig.~\ref{Fig:ind}.
In the $\phi$ mass region, the LO contribution acts as a smooth background because the process $J/\psi\to\pi^0\phi\to\pi^0\gamma^*$ is forbidden by isospin conservation. The prominent structure around 1.02~GeV is therefore generated by the $\phi$ resonance arising from the EM transition.

\begin{table}[htp]
{\footnotesize
\renewcommand{\arraystretch}{2}
\newcommand{\tabincell}[2]{\begin{tabular}{@{}#1@{}}#2\end{tabular}}
\begin{ruledtabular}
\begin{tabular}{c|cccc}
 Process &  Total & LO & EM &data~\cite{ParticleDataGroup:2024cfk} \\
\colrule
$J/\psi\to \pi^0\rho^0~(10^{-3})$  & $8.9\pm 2.0$ & $11.4\pm 1.9 $ & $0.5\pm 0.2$ & $6.2\pm 0.6$ \\
\colrule
$J/\psi\to \pi^0\omega~(10^{-4})$  & $6.6\pm 3.2$ & $0.1\pm 0.0 $ & $6.5\pm 3.2$ & $4.5\pm 0.5$\\
\colrule
$J/\psi\to \pi^0\phi~(10^{-6})$ & $6.8\pm 3.3$ & - & $6.8\pm 3.3$ & $2.94~\mathrm{or}~0.124$\\
\end{tabular}
\end{ruledtabular}
\caption{Prediction of $\mathrm{Br}[{J/\psi\to \pi^0V}]$ from our model. }
\label{tab:br2}
}
\end{table}
In the last graph of Fig.~\ref{Fig:ind}, it is evident that the LO contribution dominates in the energy region from 1.2~GeV to $M_\psi-m_{\pi^0}$. The LO contribution should be primarily driven by the $\rho'$ resonance, which is included according to Eq.~(\ref{Eq:BW}) and its coupling is absorbed into $\beta'_{\psi\pi^0\gamma^*}$.
The charmonium contribution turns larger than the EM contribution in this region but remains smaller than the LO one. Only at the very high-energy end (near 3 GeV) does the charmonium contribution approach the same order of magnitude of the LO term.
Owing to limited data and an incomplete understanding of the $J/\psi$-$J/\psi$-$\pi^0$ interaction, a precise description of $G_{\psi\pi^0\gamma^*}(q^2)$ near the charmonium threshold remains challenging. Nonetheless, our analysis shows that the charmonium part has a limited impact on the low-energy physics, which is the main focus of this work. Note that the charmonium production threshold itself is not yet reached within the plotted energy range. We anticipate that future experiments will provide improved data coverage above 1.2 GeV.


Using the parameters given in Table \ref{tab:para}, one can calculate out the branching ratios, $\mathrm{Br}[{J/\psi\to \pi^0V}]$. The results are presented in Table~\ref{tab:br2}.
As can be found, our predictions agree well with the data \cite{ParticleDataGroup:2024cfk}. To elucidate the underlying dynamics in each decay channel, we list the individual LO and EM contributions separately in the third and fourth columns.
The LO contribution dominates the $J/\psi\to \pi^0\rho^0$ decay width, while the EM one dominates that of $J/\psi\to \pi^0\omega$, consistent with our earlier discussion.
This allows us to conlcude quantitatively that the $J/\psi\to \pi^0\rho^0$ decay proceeds predominantly via the strong interaction,  whereas the $J/\psi\to \pi^0\omega$ decay originates primarily from the electromagnetic transition.
The LO contribution to the ${\rm Br}[J/\psi\to \pi^0\omega]$ is much smaller than the EM one, implying that the IB effect ($\rho$-$\omega$ mixing) is much less important than the EM transition.
The $J/\psi\to \pi^0\phi$ decay proceeds entirely via the EM transition.

\begin{figure}[htbp]
    \centering
    \includegraphics[width=1\linewidth]{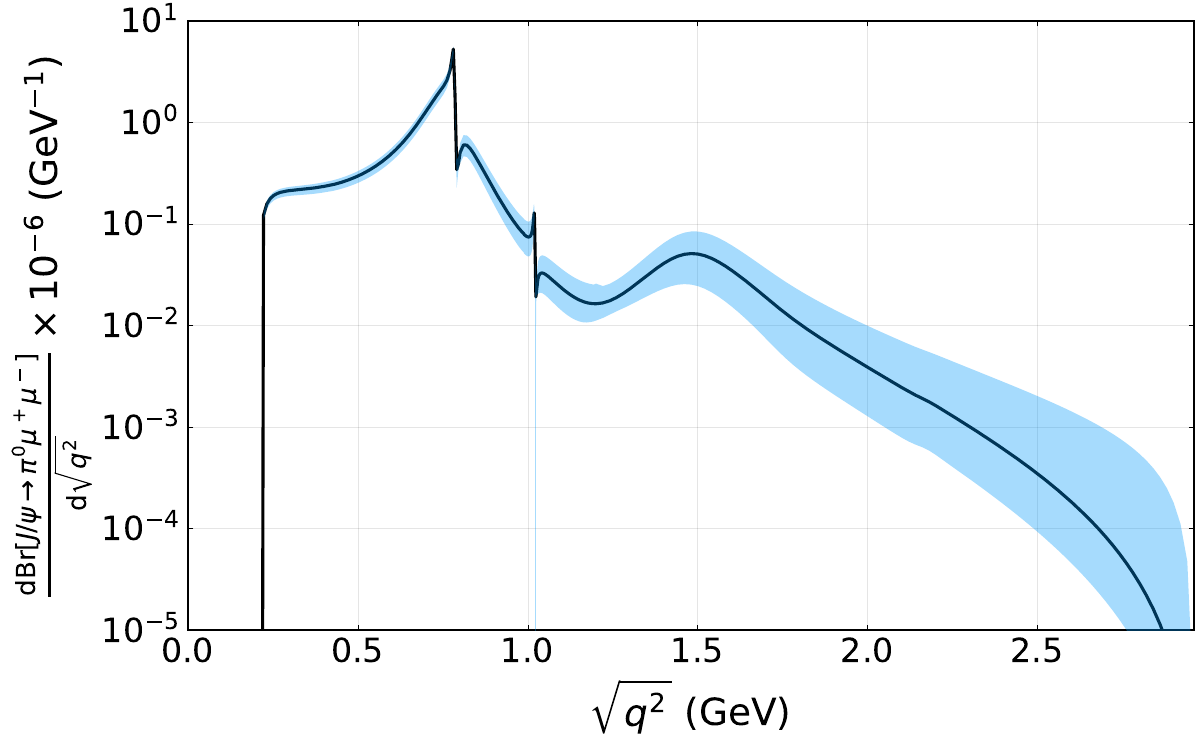}
    \caption{Predictions of $\frac{\mathrm{dBr}[{J/\psi\to\pi^0\mu^+\mu^-}]  }{\mathrm{d}\sqrt{q^2} } $.}
    \label{fig:db}
\end{figure}
The prediction of the $\mu^+\mu^-$ invariant mass spectra for  $\frac{\mathrm{dBr}[{J/\psi\to\pi^0\mu^+\mu^-}]  }{\mathrm{d}\sqrt{s} } $ is shown in Fig.~\ref{fig:db}.
Similar to the case of $J/\psi\to\pi^0e^+e^-$ decays, there are obvious structures around the $\rho,\omega,\phi$ regions. As discussed previously, these processes play a crucial role in extracting the pion TFF at $q_1^2=M_\psi^2$ as well as in testing the lepton universality.
We therefore encourage experimental measurements of these decay channels.

\section{Conclusion}\label{sec5}
In this work, we analyze the recent BESIII measurements of the $e^+e^-$ invariant mass spectrum in $J/\psi\to\pi^0e^+e^-$~\cite{BESIII:2025xjh} within the framework of resonance chiral theory.
Our model describes the data well across the entire energy region, from $2m_e$ to $M_\psi-m_{\pi^0}$.
In particular, we find that both strong interaction and electromagnetic  transition are essential for describing the physics in the low-energy region below 1.2~GeV. The structures around the $\rho$, $\omega$, and $\phi$ resonances in the invariant mass spectrum are likely generated by the interference between the strong interaction and electromagnetic transition, whereas the isospin-breaking effect (such as $\rho$-$\omega$ mixing) in the strong interaction appears to be much smaller.
We also provide predictions for the branching ratios of $J/\psi\to \pi^0V$ with $V=\rho,\omega,\phi$, which agree well with current experimental data. Quantitatively, the decay process $J/\psi\to \pi^0\rho^0$ is dominated by the strong interaction, while the other two channels, $J/\psi\to \pi^0\omega$ and $\pi^0\phi$, arise primarily from electromagnetic transitions. Future measurements of the lepton-pair invariant mass spectra in $J/\psi$ hadronic decays will be valuable for improving our understanding of the physics in the charmonium region.

\section*{Acknowledgements}
 This work is supported by the National Natural Science Foundation of China (NSFC) with Grants No.12322502, 12335002, Joint Large Scale Scientific Facility Funds of the NSFC and Chinese Academy of Sciences (CAS) under contract No.U1932110, Hunan Provincial Natural Science Foundation with Grant No.2024JJ3044, and Fundamental Research Funds for the central universities.

\bibliographystyle{unsrt}
\bibliography{ref}
\end{document}